\documentclass[
,final            
  ]
  {aipproc}

\layoutstyle{6x9}
\newcommand{\be}{\begin{eqnarray}}
\newcommand{\ben}{\begin{eqnarray}\nonumber}
\newcommand{\ee}{\end{eqnarray}}

\begin{document}

\title{Probing Metastability at the LHC}


\author{L. Clavelli}{
  address={University of Alabama\\Tuscaloosa AL 35487 USA}
}

\begin{abstract}
%
%
Current attempts to understand supersymmetry (susy) breaking are focused on the idea that we are not in the ground state of the universe but, instead, in a metastable state that will ultimately decay to an exactly susy ground state.  It is interesting to ask how experiments at the Large Hadron Collider (LHC) will shed light on the properties of this future supersymmetric universe.  In particular we ask how we can determine whether this final state has the possibility of supporting atoms and molecules in a susy background.  
\end{abstract}

\maketitle


   Although it was recognized early that susy Higgs models with an extra singlet were at best metastable \cite{Fayet} with the true ground state being exactly supersymmetric, several decades were spent trying to find a susy breaking mechanism that would allow a broken susy ground state. For reviews see \cite{Kraml,Barger}. The discovery of a non-zero vacuum energy density
\be
    \epsilon = (5.9\pm 0.2 ) {\displaystyle meV}^4 
\ee 
was the first step toward the currently dominant point of view that we live in a metastable universe destined to ultimately undergo a phase transition to a susy world.
There should be no doubt that physics will eventually be able to causally relate this milli-eV vacuum energy scale to other phenomena but, at present, our only understanding 
is based on the Anthropic Principle which observes that, within narrow bounds, the laws of physics are such as to allow the rise of intelligent life.  The string theory approach to this observation is the suggestion that there may be an enormous number of local minima in the effective potential of the universe so that the probability that some number of them would be consistent with the anthropic bound \cite{Weinberg} is naturally large.
In any case, the earliest and basic string manifestations including fermions were exactly supersymmetric so it should be expected that an exactly susy universe is among the possible worlds.
See, for example, \cite{Giddings}.  Whether one of these is the ground state or whether there are also
local minima with negative vacuum energy (collapsing universes) remains unclear.

   The existence of an inflationary era with a large vacuum energy in the very early universe suggests that there has been at least one phase transition in the past so the transition to a future susy state may not seem too unworldly.  This scenario provides an interesting alternative to the conventional cosmology.  It could be \cite{Gorsky,Voloshin} that the
transition rate is enhanced in dense matter so that the final decay of the false vacuum is presently foreshadowed by violent astrophysical events including gamma ray bursts and supernovae\cite{CK,CP}.

   We will assume, as in the Lagrangian susy Higgs models, that the exactly susy minima
with zero vacuum energy are the true ground states of the universe.  The question then arises why the universe did not begin in the exactly susy state or rapidly fall into it.
This question may ultimately have a traditional statistical mechanical answer but one could also observe that an initial exponential expansion was important for the eventual rise of life by facilitating structure formation and the escape of the baby universe from the primordial fireball.  

   From the anthropic point of view, even if a broken susy world was important at the beginning, one could ask whether life could survive or re-evolve following the ultimate phase transition to exact susy.  The properties of a future susy universe have been studied in a preliminary way \cite{future}.  The dominant feature of an exact susy world is, of course, the degeneracy of bosons and fermions which leads to a vastly diminished role for the Pauli Principle.  Some recent papers \cite{CL,CS} have suggested that ionic and covalent binding of molecules could still exist in a susy world leaving open, perhaps, the possibility that supersymmetric life forms are not ruled out. 

The susy Higgs models could provide models for the low lying part of the string landscape or, with a large number of scalar fields, could even produce many local minima.
\begin{figure}
  \includegraphics[height=.3\textheight]{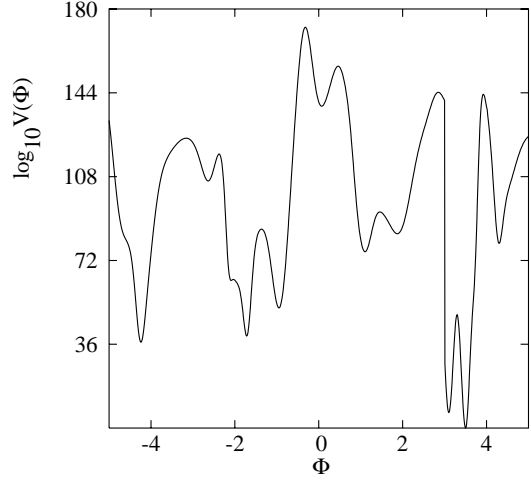}
  \caption{The effective potential in a positive definite, bi-cuspid landscape.  The vertical scale is broken to become linear at low potential.}
\end{figure}
The most general renormalizable superpotential
with a pair of Higgs doublets and an extra singlet Higgs is
\be
    W = \lambda \left( S (H_u\cdot H_d - v^2) + \frac{\lambda^\prime}{3} S^3
           + \frac{\mu_0}{2}S^2 \right) \quad .
\ee
The F terms in the scalar potential are
\be
   V_F = \lambda^2 \left( |H_u\cdot H_d - v^2 + \lambda^\prime S^2 + \mu_0 S|^2   + |S|^2 (|{H_u}|^2 + |{H_d}|^2)\right) \quad .
\ee
The overall scale of the vacuum energy and the Higgs mass squared eigenvalues is set by
$\lambda^2$ which we put equal to unity.
Of course, the introduction of dimensionful parameters, $v$ and $\mu_0$
could lead to problems analogous to the $\mu$ problem in the MSSM.
With some loss of generality, $\mu_0$ can be set to zero by imposing an
R parity invariance under $S \rightarrow -S$ in the scalar potential.
A non-zero $\mu_0$ is, however, interesting as an additional source of CP violation.  A non-zero $v$ parameter is interesting as a source of electroweak symmetry breaking in a susy ground state.  To the $F$ term in the scalar potential must be added soft terms which at present we have for simplicity restricted to soft mass terms and $D$ terms which vanish at the minima of the potential assuming equal soft mass terms for $H_u$ and $H_d$.  Our primary interest is whether or not $v$ is non-zero so it is sufficient at present to set these soft masses equal.
\be
    V = V_F + m_S^2 |S|^2 + m_H^2 (|H_u|^2+|H_d|^2) \quad .
\ee
The critical points \cite{critical} where all derivatives vanish are
determined by
\be
     \frac{\partial V}{\partial S^*}|_{_0}=
0 = (2 \lambda^\prime S_0^* + \mu_0^*) (v_0^2-v^2 + \mu_0 S_0 + \lambda^\prime S_0^2) 
+ S_0 (2 |{v_0}|^2 + m_S^2)
\ee
and
\be
     \frac{\partial V}{\partial H_u^*}|_{_0} =
0 = v_0^* (v_0^2-v^2 + \mu_0 S_0 + \lambda^\prime  S_0^2 ) + v_0 (|{S_0}|^2 + m_H^2) \quad .
\ee
The soft masses depend on the susy breaking mechanism and vanish in the transition to exact susy. 
If $v$ is non-zero, degenerate exact susy ground states are found in \cite{critical}, 
one of which has electroweak symmetry breaking. 
In addition two susy breaking critical points are found which, in the limit of vanishing soft masses are saddle points and not true minima.
We have suggested that these could be promoted to true local minima by ``soft breaking terms'' which, in turn, could be induced by dynamical susy breaking in some hidden sector.  We have, up to now, only considered soft mass terms.

The value of the Higgs potential at the broken susy mininum (solution 4 in \cite{critical}) is
\ben
  V_4(0)=  |S_0|^4 + 2|S_0 v_0|^2 + m_H^4 + m_S^2|S_0|^2 + 2 m_H^2 (|v_0|^2+|S_0|^2) \quad .
\quad . 
\ee
One can choose the soft masses to make the vacuum energy in solution 4 agree with the dark energy observation: $(5.9\pm0.2)$meV$^4$.
(At least one of $m_S^2$ and $m_H^2$ must be negative which is allowable as long as the physical eigenstates are positive.)  Of course, this requires a fine tuning and does not explain the low vacuum energy.  Other sources of dark energy such as thermal effects and compactification effects would also come into play and, in the absence of interference, would also have to be small.  Nevertheless, this moves the problem into the hidden sector and provides an example of the Higgs vacuum energy being small compared to the natural scale of $|v_0|=176$ GeV. Models of 
dynamical susy breaking in a hidden sector have been presented 
\cite{ISS,Dienes} although the mediation mechanism and the quantitative connection to soft mass terms remains a subject for further work.

Only if the parameter $v$ is found to be greater than zero is there a 
susy ground state with electroweak symmetry breaking allowing atomic and molecular binding and, therefore, the possibility of susy life forms. 
Such a non-zero $v$ parameter is characteristic of the ``nearly minimal 
susy standard model'', (nMSSM) but not of the minimal (MSSM) or next to minimal models (NMSSM).  This parameter, like all the parameters in the Lagrangian, can be determined by sufficiently detailed experimentation at a high energy collider such as the LHC
\cite{BLS} but this will not be a simple job since it requires measuring the vacuum expectation value of the scalar Higgs and the other parameters of the Higgs potential. 
For example, one can test for a non-zero $v$ by measuring the Higgs mixing term after the other parameters, $\lambda^\prime$, $\mu_0$, and $S_0$ have been determined.  In the
most general model this term is
\be
 V_{mixing} = H_u \cdot H_d \left(v_0^2 - v^2 + \lambda^\prime S_0^2 + \mu_0 S_0\right)^\dagger + h.c.  \quad .
\ee
This relation, of course, receives corrections proportional to electroweak gauge couplings, which may be small, and to
differences of soft doublet Higgs masses.

If a non-zero $v$ is found, it will be of interest to determine whether it is less than the Higgs doublet expectation value $v_0$ in the broken susy state in which case there could be an exothermic transition in dense matter from the broken susy universe to the exact susy phase. In this case, if a susy bubble overtakes a matter filled region,
the normal matter will be converted to susy without the input of extra energy.

In \cite{critical} we have explored correlations between the various
parameters in the two cases of $v < |v_0|$ and $v > |v_0|$.
An interesting property of the model with negative soft masses squared is that the coupling constants can be in the perturbative region.  In the second of \cite{critical} we have 
considered phases in the potential which can be related to CP violating phenomena.


\begin{theacknowledgments}
  We acknowledge helpful discussions with Roman Nevzorov, with Zachary Burell, and with Irina Perevalova.
\end{theacknowledgments}

\end{document}